# Thermal conductance through molecular wires

Dvira Segal and Abraham Nitzan

School of Chemistry, Tel Aviv University, Tel Aviv, 69978, Israel

and

Peter Hänggi

Institut für Physik, Universität Augsburg, Universitätsstr. 1, D-86135 Augsburg, Germany

**Abstract**

We consider phononic heat transport through molecular chains connecting two thermal reservoirs. For relatively short molecules at normal temperatures we find, using classical stochastic simulations, that heat conduction is dominated by the harmonic part of the molecular force-field. We develop a general theory for the heat conduction through harmonic chains in 3-dimensions. Our approach uses the standard formalism that leads to the generalized Langevin equation for a system coupled to a harmonic heat bath, however the driving and relaxation terms are considered separately in a way that leads directly to the steady state response and the heat current under non-equilibrium driving. A Landauer-type expression for the heat conduction is obtained, in agreement with other recent studies. We used this general formalism to study the heat conduction properties of alkane. We find that for relatively short (1-30 carbon molecules) the length and temperature dependence of the molecular heat conduction results from the balance of three factors: (i) The molecular frequency spectrum in relation to the frequency cutoff of the thermal reservoirs, (ii) the degree of localization of the molecular normal modes and (iii) the molecule-heat reservoirs coupling. The fact that molecular modes at different frequency regimes have different localization properties gives rise to intricate dependence of the heat conduction on molecular length at different temperature. For example, the heat conduction increases with molecular length for short molecular chains at low temperatures. Isotopically substituted disordered chains are also studied and their behavior can be traced to the above factors together with the increased mode localization in disordered chain and the increase in the density of low frequency modes associated with heavier mass substitution. Finally, we compare the heat conduction obtained from this microscopic calculation to that estimated by considering the molecule as a cylinder characterized by a macroscopic heat conduction typical to organic solids. We find that this classical model overestimates the heat conduction of single alkane molecules by about an order of magnitude at room temperature. Implications of the present study to the problem of heating in electrically conducting molecular junctions are pointed out.

# 1. Introduction

The investigation of the electrical conductance of nanowires is in the focus of the quest for developing novel submicron and nano-size electrical devices. Molecular devices already demonstrated include molecular wires, field effect transistors, single electron transistors, molecular diodes, rectifiers and switches.[1,2] Localized Joule heating poses a crucial question over the functionality and reliability of such devices. The combination of small molecular heat capacity and inefficient heat transfer away from it may cause a large temperature increase that will affect the stability and integrity of the molecular junction. The rate at which heat is transported away from the conducting junction is therefore crucial to the successful realization of nano electronics devices.

As in macroscopic solids conductors heat can be carried away from the junction by electrons and phonons. In metals heat flow is dominated by electrons, while in insulators heat is transmitted solely by phonons. This study focuses on the phononic mode of heat transfer. Theoretical interest in this mode of heat transfer in solids goes back to Peierls' early work.[3] Recently it was found that thermal transport properties of nanowires can be very different from the corresponding bulk properties. For example, Rego and Kirczenow[4] have shown theoretically that in the low temperature ballistic regime, the phonon thermal conductance of a 1 dimensional quantum wire is quantized, and have obtained $g = \pi^2 k_B^2 T / 3h$ as the universal quantum conductance unit, where $k_B$ and $h$ are the Boltzmann and Planck constants, respectively, and $T$ is the temperature. Also of considerable interest are attempts to derive the macroscopic Fourier law of heat conduction in 1-dimensional systems from microscopic considerations. The Fourier law is a relationship between the heat current $J$ per unit area $\mathcal{A}$ and the temperature gradient $\nabla T$

$$J / \mathcal{A} = -\tilde{K} \nabla T \tag{1}$$

where $\mathcal{A}$ is the cross-section area normal to the direction of heat propagation and $\tilde{K}$ is the thermal conductivity (the thermal conductance $K$ is defines as $K = J / \Delta T$). Perfect harmonic chains were theoretically investigated by Rieder and Lebowitz[5] and by Zürcher and Talkner[6] who found that heat flux in these systems is proportional to the temperature difference and not to the temperature gradient. Consequently, the thermal conductivity diverges with increasing chain length. Anomalous heat conduction was also found in 1-dimensional models of colliding hard particles.[7,8] Different models that potentially avoid this divergence and yield Fourier law conduction were discussed. Some invoke impurities and disorder[9,10], others[11,12] consider anharmonicity as the source of normal heat conduction. Numerical simulations for chains with a random potential were performed by Mokross,[13] and the role of phonon-lattice interaction was studies by Hu et al.[14] Still, there is yet no convincing and conclusive result about the validity of Fourier law in 1D systems.

Experimentally, remarkable progress has been achieved in the last decade in nanoscale thermometry, and measurements on the scale of the mean free path of phonons and electrons are possible. Using scanning thermal microscopy methods one can obtain the spatial temperature distribution of the sample surface, study local thermal properties of materials, and perform calorimetry at nanometric scale.[15,16] The thermal conductivity and thermoelectric power of single carbon nanotubes were studied both experimentally[17] and theoretically[18,19]. In a different experiment, Schwab et al[20] have observed the quantum thermal conductance in a nano fabricated 1D structure, which behaves essentially like a phonon waveguide. Their results agree with the theoretical predictions.[4] These and other experimental and theoretical developments in this field have been recently reviewed.[21]

In a recent paper[22] we have estimated the rate of heat generation in a model of a current carrying molecular junction. We have found that a substantial (~0.1-0.5) fraction of the voltage drop across the junction is dissipated as heat on the molecule, implying that a power of the order of $10^{11}$ eV/s may be released as heat on a molecular bridge carrying a current of 10nA under a bias of 1eV. This would cause a substantial temperature rise in the molecule unless heat is effectively carried into the metal leads. This motivates a study of molecular heat conduction. In Ref. 22 we have used a simple classical continuum model (Fig. 1) in which the molecular bridge is represented by a cylinder characterized by a heat conduction coefficient ~ $\sigma_h = 10^{-4}$ cal/(s·cm·K), typical to solid saturated alkanes. However classical heat conduction theory is expected to overestimate the heat flux through a single molecule that has a discrete vibrational frequency spectrum, and a molecular level treatment is needed for a correct description of this process.

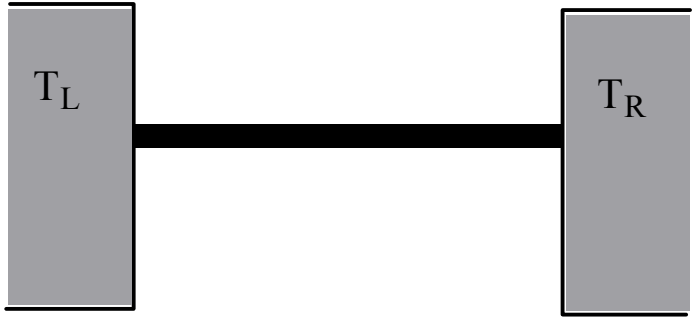

FIG. 1: A schematic representation of a molecular chain connecting two heat reservoirs.

In the present paper we address this problem, focusing on the steady state phononic heat transfer through a molecule connecting two macroscopic thermal reservoirs of different temperatures. The calculation is based on the generalized Langevin equation (GLE) approach[23,24,25,6] which is recast for a model of single molecule junction. The resulting expression for the heat current through harmonic molecules is analogous to the Landauer expression for electrical conductance.[26] We apply our formalism to realistic models of molecular systems: alkane chains of varying lengths, using the HyperChem package to generate molecular structures and obtain their vibrational (normal mode) spectrum, and using a Debye model for the thermal reservoirs. This enables us to study the dependence of the heat transfer on the bridge length, the temperature and molecular composition, as well as on the spectral properties of the reservoirs.

Section 2 introduces our formal model for phononic heat transfer through molecular bridges, and Section 3 describes our theoretical approach that yields Langevin-type equations of motion for the molecular subsystem and an expression for the heat transfer rate. Section 4 presents some numerical results for the heat conduction by alkane and alkane-like bridges connecting Debye solids, and discusses their implications. Section 5 concludes.

## 2. Model

We consider a molecule connecting two independent macroscopic solids, L and R, which are held at fixed temperatures $T_L$ and $T_R$ respectively. In steady state there is a constant heat flow between these two heat reservoirs through the molecule. A schematic representation of the model is depicted in Figure 1. The Hamiltonian of this system is a sum of the molecular Hamiltonian, $H_M$, the Hamiltonian of the solid baths, $H_B$, and the molecule-bath interaction $H_{MB}$,

$$H = H_M + H_B + H_{MB} \qquad (2)$$

Even though the heat flow in our system is 1-dimensional, the thermal reservoirs and the molecular bridge are three dimensional objects. The reservoirs are represented as systems of independent harmonic oscillators at thermal equilibrium. In what follows we consider a harmonic molecule coupled linearly to these thermal environments. Anharmonic effects will be discussed in Section 4 where we show that for the relative short chains considered here and at room temperature they are relatively small. For simplicity we assume that only the end atoms, $i=1$ and $i=N$, of the molecular chain are coupled (linearly) to the solids. To simplify our presentation we write the molecule-bath coupling in 1-dimension (the analogous 3-dimensional expressions that are used in the computations are given in Appendix A). The Hamiltonian associated with the environment and its coupling to the molecule is then given by[25]

$$H_B + H_{MB} = \sum_l \left\{ \frac{1}{2} m_l \omega_l^2 \left( x_l - \frac{g_{1,l} x_1}{m_l \omega_l^2} \right)^2 + \frac{p_l^2}{2 m_l} \right\} + \sum_r \left\{ \frac{1}{2} m_r \omega_r^2 \left( x_r - \frac{g_{N,r} x_N}{m_r \omega_r^2} \right)^2 + \frac{p_r^2}{2 m_r} \right\}$$
$$= H_L + H_R + H_{MB} \qquad (3)$$

where

$$H_L = \sum_l \left\{ \frac{1}{2} m_l \omega_l^2 x_l^2 + \frac{p_l^2}{2 m_l} \right\} \quad ; \quad H_R = \sum_r \left\{ \frac{1}{2} m_r \omega_r^2 x_r^2 + \frac{p_r^2}{2 m_r} \right\} \qquad (4)$$

$$H_{MB} = H_{ML} + H_{MR} = \sum_l \frac{1}{2} \frac{g_{1,l}^2 x_1^2}{m_l \omega_l^2} - \sum_l g_{1,l} x_1 x_l + \sum_r \frac{1}{2} \frac{g_{N,r}^2 x_N^2}{m_r \omega_r^2} - \sum_r g_{N,r} x_N x_r \qquad (5)$$

and where $x_j$, $p_j$, $m_j$ and $\omega_j$ ($j=l,r$) are coordinates, momenta, masses and frequencies associated with the degrees of freedom of the reservoirs, and where the subscripts $l$ and $r$ are used for the left (L) and right R reservoirs, respectively. The molecule-solids coupling is characterized by the constants $g_{1,l}$ and $g_{N,r}$, and $x_1$ and $x_N$ are the coordinates of the molecule end atoms.

In what follows we use a generic description of the molecular bridge, representing it as a set of $N$ independent- collective harmonic oscillators

$$H_M = \sum_{k=1}^{N} \left\{ \frac{1}{2} \omega_k^2 \bar{x}_k^2 + \frac{\bar{p}_k^2}{2} \right\} \qquad (6)$$

where $\bar{x}_k$ and $\bar{p}_k$ are the (mass weighted) displacement and momentum associated with the normal mode $k$. The normal mode representation (6) is obtained from the atomic (local) coordinate representation by the standard procedure of first transforming the local coordinates $x_i$ and $p_i$ ($i=1,…,N$) into mass weighted coordinates $x_i\sqrt{m_i} \to x_i$ and $p_i/\sqrt{m_i} \to p_i$, then diagonalizing the molecular Hessian matrix. This defines a linear transformation

$$\mathbf{x} = \mathbf{C}\bar{\mathbf{x}} \qquad (7)$$

that relates the $N$-vector $\mathbf{x}$ of mass weighted local coordinates to the $N$-vector $\bar{\mathbf{x}}$ of molecular normal modes. The $N \times N$ matrix $\mathbf{C}$ is assumed in what follows to be known. Note that the coordinates $x_1$ and $x_N$ that appear in (3) and (5) are local, corresponding to the positions of the first and last atoms in the linear molecular chain, and when re-expressed in terms of the molecular normal modes results in coupling terms that connect *all normal modes* to the thermal reservoirs. A similar transformation to mass weighted representation is done also on the normal modes of the macroscopic solids. The Hamiltonian terms (3)-(5) then become

$$H_L = \sum_l \left\{ \frac{1}{2}\omega_l^2 \bar{x}_l^2 + \frac{\bar{p}_l^2}{2} \right\} \; ; \qquad H_R = \sum_r \left\{ \frac{1}{2}\omega_r^2 \bar{x}_r^2 + \frac{\bar{p}_r^2}{2} \right\} \qquad (8)$$

$$H_{MB} = \sum_{l,k,k'} \frac{1}{2\omega_l^2} V_{l,k} V_{l,k'} \bar{x}_k \bar{x}_{k'} - \sum_{l,k} V_{l,k} \bar{x}_l \bar{x}_k + \sum_{r,k,k'} \frac{1}{2\omega_r^2} V_{r,k} V_{r,k'} \bar{x}_k \bar{x}_{k'} - \sum_{r,k} V_{r,k} \bar{x}_r \bar{x}_k \qquad (9)$$

where the mass weighted bath coordinates are denoted by $\bar{x}$ and $\bar{p}$. In Eq. (9), the sums over $k$ and $k'$ go over the $N$ molecular normal modes, while the indices $l$ and $r$ denote, as before, the modes of the L and R solids, respectively. The transformed molecule–baths coupling constants are given by

$$\begin{aligned} V_{l,k} &\equiv V_l C_{1,k} & V_l &= \frac{g_{1,l}}{\sqrt{m_1 m_l}} \\ V_{r,k} &\equiv V_r C_{N,k} & V_r &= \frac{g_{N,r}}{\sqrt{m_N m_r}} \end{aligned} \qquad (10)$$

The total Hamiltonian is the sum of the terms in Eqs. (6), (8) and (9). In this representation all the molecular information is contained in its normal modes frequencies, the transformation matrix $\mathbf{C}$ and the coupling constants $V_l$ and $V_r$. It should be evident that the same treatment can be done for three dimensional molecule-bath coupling (see Appendix A).

In the following section we use this harmonic model to calculate the heat transport properties of molecular junctions. This assumes that heat transport in such junctions is dominated by the harmonic part of the molecular nuclear potential. The extent to which this assumption holds will be examined later.

## 3. Calculation of the steady state heat flux

Here we use the model outlined in Sect. 2 to calculate the steady state phonon energy transfer between the two thermal baths through the molecular link. Starting from the coupled classical equations of motion for all (molecular and baths) modes, we derive a set of classical Langevin equations for the molecular modes by projecting out the baths degrees of freedom. Then, using the classical equations as a guide, we construct the corresponding quantum Langevin equations for the molecular system. Transformation to the frequency domain makes it possible to extract steady state information and finally yields the steady-state heat current from the transformed equations.

*Equations of motion*. The classical equations of motion for all modes are obtained from the Hamilton's equations $-\partial H / \partial \bar{q}_i = \dot{\bar{p}}_i$ ; $\partial H / \partial \bar{p}_i = \dot{\bar{q}}_i$. Here $H$ is the sum of Eqs. (6), (8) and (9). This leads after rearrangement to

$$\ddot{\bar{x}}_k = -\omega_k^2 \bar{x}_k + \sum_l V_{l,k} \bar{x}_l + \sum_r V_{r,k} \bar{x}_r - \sum_{l,k'} \frac{1}{\omega_l^2} V_{l,k} V_{l,k'} \bar{x}_{k'} - \sum_{r,k'} \frac{1}{\omega_r^2} V_{r,k} V_{r,k'} \bar{x}_{k'}, \quad (11)$$

$$\ddot{\bar{x}}_l = -\omega_l^2 \bar{x}_l + \sum_k V_{l,k} \bar{x}_k$$

$$\ddot{\bar{x}}_r = -\omega_r^2 \bar{x}_r + \sum_k V_{r,k} \bar{x}_k \quad (12)$$

Note that different molecular modes are coupled to each other through their interaction with the baths.

*Langevin equations*. Next we follow a standard procedure[25] in which Eqs. (12) are formally integrated and used in (11) to yield a set of generalized Langevin equations for the molecular modes. In the resulting equations the effect of the thermal environments appears in new driving forces and damping terms. This procedure (Appendix B) leads to

$$\ddot{\bar{x}}_k = -\omega_k^2 \bar{x}_k + \sum_l V_{l,k} \tilde{x}_l + \sum_r V_{r,k} \tilde{x}_r$$
$$- \sum_{l,k'} \frac{V_{l,k} V_{l,k'}}{\omega_l^2} \int_{t_0}^t \dot{\bar{x}}_{k'}(\tau) \cos(\omega_l(t-\tau)) d\tau - \sum_{r,k'} \frac{V_{r,k} V_{r,k'}}{\omega_r^2} \int_{t_0}^t \dot{\bar{x}}_{k'}(\tau) \cos(\omega_r(t-\tau)) d\tau \quad (13)$$

where $\tilde{x}_l$ and $\tilde{x}_r$ evolve according to

$$\ddot{\tilde{x}}_l = -\omega_l^2 \tilde{x}_l \quad \text{and} \quad \ddot{\tilde{x}}_r = -\omega_r^2 \tilde{x}_r \quad (14a)$$

or

$$\tilde{x}_l(t) = \tilde{x}_l(t_0) \cos(\omega_l(t-t_0)) + \frac{\dot{\tilde{x}}_l(t_0)}{\omega_l} \sin(\omega_l(t-t_0)) \quad (14b)$$

and similarly for the *r* modes.

Equation (13) is a generalized Langevin equation for the molecular mode *k*. The terms

$$M_L(t) \equiv \sum_{l,k'} \frac{V_{l,k} V_{l,k'}}{\omega_l^2} \int_{t_0}^t \dot{\bar{x}}_{k'}(\tau) \cos(\omega_l(t-\tau)) d\tau = \sum_{k'} \int_{-\infty}^t \dot{\bar{x}}_{k'}(\tau) \gamma_{k,k'}^L(t-\tau) d\tau \quad (15)$$

(we take $t_0 \to -\infty$ because we are interested in the long-time, steady state situation) with the memory kernel or time dependent friction

$$\gamma_{k,k'}^L(t) = \sum_l \frac{V_{l,k} V_{l,k'}}{\omega_l^2} \cos(\omega_l t) \quad (16)$$

and the similar terms with *R* and *r* replacing *L* and *l* are the damping terms that result from eliminating the degrees of freedom of the L and R baths. The corresponding "random forces" are

$$F_L^{(k)}(t) = \sum_l V_{l,k} \tilde{x}_l \quad ; \quad F_R^{(k)}(t) = \sum_r V_{r,k} \tilde{x}_r \quad (17)$$

Their random character follows from the random distribution of the initial conditions in Eq (14). These random forces and memory kernels are related to each other by a fluctuation-dissipation type relation.

$$\left\langle F_L^{(k)}(t) F_L^{(k')}(0) \right\rangle = \sum_{l,l'} V_{l,k} V_{l',k'} \left\langle \left[ \tilde{x}_l(0) \cos(\omega_l t) + \frac{\dot{\tilde{x}}_l(0)}{\omega_l} \sin(\omega_l t) \right] \tilde{x}_{l'}(0) \right\rangle \quad (18)$$

Using the classical relationships $\left\langle \tilde{x}_l(0) \tilde{x}_{l'}(0) \right\rangle = \delta_{l,l'} k_B T_L / \omega_l^2$ and $\left\langle \dot{\tilde{x}}_l(0) \tilde{x}_{l'}(0) \right\rangle = 0$ this becomes $\left\langle F_L^{(k)}(t) F_L^{(k')}(0) \right\rangle = k_B T_L \gamma_{k,k'}^L(t)$. Similar relations hold for the *R* quantities.

The above procedure is a standard derivation of a generalized Langevin equation usually used to describe a system coupled to its thermal environment. In our case, when the system is driven by different environments out of equilibrium with each other, it is useful to look at the resulting equations as describing a driven system. To this end we note that Eqs. (13) and (14), viewed as a set of deterministic linear equations, describe a system $\{x_k\,;\,k=1,...,N\}$ of damped harmonic oscillators, driven by a set of oscillators $\{x_j\,;\,j\in L,R\}$ that move independently of the driven system (in our case – according to Eqs. (14) with initial conditions that will be averaged on at the end of the calculation). These oscillators act on the system additively, and the effect of each may be considered separately. Our following derivation is facilitated by considering a version of Eq. (13) with only one driving mode $x_0$ of frequency $\omega_0$,

$$\ddot{\bar{x}}_k = -\omega_k^2 \bar{x}_k + V_{0,k}\tilde{x}_0 - \sum_{k'} \int_{-\infty}^{t} (\gamma_{k,k'}^L(t-\tau) + \gamma_{k,k'}^R(t-\tau))\dot{\bar{x}}_{k'}(\tau)d\tau\,;\ k=1,...,N \quad (19)$$

$$\tilde{x}_0(t) = \tilde{x}_0(t_0)\cos(\omega_0(t-t_0)) + \frac{\dot{\tilde{x}}_0(t_0)}{\omega_0}\sin(\omega_0(t-t_0)) \quad (20)$$

At long time a system described by these equations approaches a steady state in which the external mode 0, which may belong to either the L or the R bath, drives all other system modes to oscillate at frequency $\omega_0$ with amplitude derived from that of the driving mode.

The formulation of our problem in terms of Eqs. (19) and (20), with Eq. (20) representing one of the external bath modes that drives the molecular system, makes it possible to address the system in non equilibrium situations. For example, Eqs. (19)-(20) describe the physics of a system in which only external mode 0 is excited while the others are at $T=0$. Moreover, the motion of mode 0, determined by the choice of $\tilde{x}_0(0)$ and $\dot{\tilde{x}}(0)$ does not have to be thermal. Furthermore, *this formulation makes it possible to calculate the flux distribution into the different bath modes given that the mode* 0 *drives the system.* To do this one needs to use the solution of Eq. (19) (obtained under the driving (20)) in Eqs. (12) to find the response of other bath modes to the driving by mode 0. Such a calculation is facilitated by replacing Eqs. (12) by their damped analogs:

$$\ddot{\bar{x}}_l = -\omega_l^2 \bar{x}_l + \sum_k V_{l,k} \bar{x}_k - \eta \dot{\bar{x}}_l \; ; \quad l \in L$$
$$\ddot{\bar{x}}_r = -\omega_r^2 \bar{x}_r + \sum_k V_{r,k} \bar{x}_k - \eta \dot{\bar{x}}_r ; \quad r \in R \qquad \eta \to 0+ \qquad (21)$$

The long time solution to Eqs. (19)-(21) is a steady state in which energy flows from the driving mode 0 into the $\{l\}$ and $\{r\}$ modes through the molecular modes $\{k\}$. In particular, the steady state heat flux channeled through, e.g., the mode $r$ is given by the rate of energy dissipation out of this mode

$$J_{0\to r} = \eta \langle \dot{\bar{x}}_r^2 \rangle_t \qquad (22)$$

The integrated fluxes, $J_{0\to L} = \sum_l J_{0\to l}$ and $J_{0\to R} = \sum_r J_{0\to r}$ from the driving mode 0 into the left and right baths should not depend on $\eta$. For the case where mode 0 belongs to the left bath, $J_{0\to R}$ and $J_{0\to L}$ correspond to the transmitted and reflected fluxes, respectively.[27]

The above formulation portrays in a somewhat new light the familiar double role, driving and damping, assumed by bath modes in such problems. In equilibrium these two actions are balanced by the fluctuation-dissipation theorem. In non-equilibrium situations it is sometime useful to consider these two roles separately. Indeed, later below we will calculate the energy (heat) flux induced by one driving mode throughout the system. The net heat flux at frequency $\omega_0$ is obtained as the difference between such fluxes originated in the two baths and weighted by the corresponding density of modes. The total heat flux is obtained by integrating over all frequencies. Before that, however, we construct the quantum equations of motion equivalent to (19)-(20).

*Quantum equations of motion.* For a system of harmonic oscillators the equations representing the classical dynamics, Eqs. (11)-(20), may be also viewed as quantum EOMs for the Heisenberg position and momenta operators. The formal connection is made as usual by first defining linear transformation on the position and momentum variables

$$\bar{x}_j(t) = \sqrt{\frac{1}{2\omega_j}} \left( a_j^*(t) + a_j(t) \right) \; ; \quad \bar{p}_j(t) = i\sqrt{\frac{\omega_j}{2}} \left( a_j^*(t) - a_j(t) \right) \; ; j=0,\{k\},\{l\},\{r\} \quad (23)$$

where we use $\hbar = 1$. Eq. (20) then yields

$$a_0(t) = a_0 e^{-i\omega_0 t} ; \qquad a_0^*(t) = a_0^* e^{i\omega_0 t}, \qquad (24)$$

where $a_0$ is the classical complex initial amplitude of the driving mode, and Eq. (19) is equivalent to the following coupled equations

$$\frac{d^2}{dt^2}\left(a_k^*(t)+a_k(t)\right) = -\omega_k^2\left(a_k^*(t)+a_k(t)\right)+V_{0,k}\sqrt{\frac{\omega_k}{\omega_0}}\left(a_0^*(t)+a_0(t)\right)$$
$$-i\sum_{k'}\sqrt{\omega_k\omega_{k'}}\int_{-\infty}^{t}d\tau\left[\gamma_{k,k'}^L(t-\tau)+\gamma_{k,k'}^R(t-\tau)\right]\left(a_{k'}^*(\tau)-a_{k'}(\tau)\right) \quad (25)$$

$$\frac{d}{dt}\left(a_k^*+a_k\right)=i\omega_k(a_k^*-a_k) \quad (26)$$

Eqs. (25) and (26) constitute the classical EOMs for the variables $a_k, a_k^*$ defined by (23). Quantization is now achieved by replacing $a_j^*$ by $a_j^\dagger$ ($j=0, \{k\}$) and regarding Eqs. (25) and (26) as equations of motions for the Heisenberg representation of the creation and annihilation operators $a_j(t)$ and $a_j^\dagger(t)$. The thermal information then enters via

$$<a_0^\dagger a_0>_L = n_L(\omega_0) = \left(e^{\beta_L\omega_0}-1\right)^{-1}; \quad <a_0^\dagger a_0>_R = n_R(\omega_0) = \left(e^{\beta_R\omega_0}-1\right)^{-1} \quad (27)$$

where $\beta = 1/k_B T$.

In what follows we will also require the quantum equations of motions for the bath modes. Using Eq. (23) in (21) we obtain[29]

$$\dot{a}_r^\dagger = i\omega_r a_r^\dagger - i\sum_k \frac{V_{r,k}}{2\sqrt{\omega_k\omega_r}}\left(a_k^\dagger+a_k\right)-(\eta/2)\left(a_r^\dagger-a_r\right) \quad (28)$$

and its complex conjugate, and similar equations for the $l$ modes.

*Frequency domain equations.* Because our system is linear, at steady-state all the modes oscillate with the driving frequency $\omega_0$. Accordingly we seek a solution of the form

$$\left(a_j^\dagger+a_j\right)=A_j e^{i\omega_0 t}+B_j e^{-i\omega_0 t}; \quad j\in\{k\},\{l\},\{r\} \quad (29)$$

which has to satisfy $B_j = A_j^\dagger$. Also, the need to satisfy $(d/dt)\left(a_j^\dagger+a_j\right)=i\omega_j(a_j^\dagger-a_j)$ (same as Eq. (26)) implies

$$\left(a_j^\dagger-a_j\right)=\left(A_j e^{i\omega_0 t}-B_j e^{-i\omega_0 t}\right)\frac{\omega_0}{\omega_j}; \quad j\in\{k\},\{l\},\{r\} \quad (30)$$

or

$$a_j^\dagger = \frac{A_j}{2} e^{i\omega_0 t}\left(1+\frac{\omega_0}{\omega_j}\right) + \frac{B_j}{2} e^{-i\omega_0 t}\left(1-\frac{\omega_0}{\omega_j}\right)$$

$$a_j = \frac{B_j}{2} e^{-i\omega_0 t}\left(1+\frac{\omega_0}{\omega_j}\right) + \frac{A_j}{2} e^{i\omega_0 t}\left(1-\frac{\omega_0}{\omega_j}\right) \quad ; \quad j \in \{k\},\{l\},\{r\} \tag{31}$$

Note that Eq. (31), taken with $j=0$, is consistent with Eq. (24). Inserting Eqs. (29) and (30), for $j=k$, into (25) and equating separately the coefficients of $e^{i\omega_0 t}$ and $e^{-i\omega_0 t}$ leads to

$$-\omega_0^2 A_k = -\omega_k^2 A_k + V_{0,k}\sqrt{\frac{\omega_k}{\omega_0}} a_0^\dagger - i\sum_{k'}\sqrt{\omega_k \omega_{k'}}\frac{\omega_0}{\omega_{k'}} A_{k'} \int_0^\infty d\tau e^{-i\omega_0 \tau}\left[\gamma_{k,k'}^L(\tau) + \gamma_{k,k'}^R(\tau)\right] \tag{32}$$

or

$$\left(\omega_k^2 - \omega_0^2 + i\omega_0\left[\gamma_{k,k}^L(\omega_0) + \gamma_{k,k}^R(\omega_0)\right]\right)A_k(\omega_0) +$$
$$+ i\omega_0 \sum_{k'\neq k}\sqrt{\frac{\omega_k}{\omega_{k'}}}(\gamma_{k,k'}^L(\omega_0) + \gamma_{k,k'}^R(\omega_0))A_{k'}(\omega_0) = \sqrt{\frac{\omega_k}{\omega_0}} V_{0,k} a_0^\dagger \tag{33}$$

where

$$\gamma_{k,k'}^L(\omega) = \int_0^\infty e^{-i\omega t}\gamma_{k,k'}^L(t)dt = \sum_l \frac{V_{l,k}V_{l,k'}}{2\omega_l^2}\left\{\pi\delta(\omega_l - \omega) + iP\left(\frac{1}{\omega_l - \omega}\right)\right\} \tag{34}$$

and a similar expression for $\gamma_{k,k'}^R(\omega)$, where P denotes principal part. To obtain (34) we have used (16) and have disregarded terms containing $(\omega + \omega_l)^{-1}$ factors. For simplicity we further invoke a standard approximation by which we disregard the imaginary part of $\gamma$ (i.e. terms affecting small frequency shifts), representing it by its real part

$$\gamma_{k,k'}^L(\omega) = \frac{\pi}{2}\frac{V_{l,k}V_{l,k'}\big|_{\omega_l=\omega}}{\omega^2}\rho_L(\omega) \tag{35}$$

For future reference we also rewrite this function, using Eq. (10), in the form

$$\gamma_{k,k'}^L(\omega) = \gamma_L(\omega)C_{1,k}C_{1,k'} \tag{36}$$

where $\gamma_L(\omega)$ is defined from this expression. Again, an equivalent expression defines $\gamma_R(\omega)$.

Eq. (33) can be solved to yield the set of operators $\{A_k\}$ associated with the Heisenberg operators for the molecular bridge modes according to Eq. (29). In a similar way, the amplitudes $\{A_l\}$ and $\{A_r\}$ associated with the bath modes according to Eq. (29)

can be obtained. For this purpose we use Eqs. (29)-(31) in (28) and again consider separately coefficients of $\exp(i\omega_0 t)$ and of $\exp(-i\omega_0 t)$. This leads to

$$A_r = \frac{\omega_r}{\left(\omega_r^2 - \omega_0^2 + i\eta\omega_0\right)} \sum_k \frac{V_{r,k}}{\sqrt{\omega_r \omega_k}} A_k; \qquad B_r = A_r^\dagger \qquad (37)$$

with $k$ going over all bridge modes. A similar equation is obtained for the operators $A_l$ of the left-side bath.

*Calculation of the heat flux*. Eqs. (37) and (33) lead to linear relationships between the operators $A_l$ (or $A_r$) that describe the driven outgoing bath modes and between the operators $a_0$ and $a_0^\dagger$ that describe the driving mode. This can be used to compute the heat flux through a system subjected to such driving. Different approaches to calculating the heat flux through a system of linear oscillators can be found in the literature[30,31,6,32,33] and a common method suitable in particular to 1-dimensional systems is based on calculating the work done by atom $i$ on its neighbor $i$-1.[6] For our model this leads to, e.g. at the right side metal-bath contact

$$J = \frac{g_{Nr}}{2m_r} \langle x_N p_r + p_r x_N \rangle. \qquad (38)$$

where the coupling parameters $g$ were introduced in Eq. (3) and the symmetrized form is needed for quantum mechanical calculations.[34] Obviously, at steady state the heat flux is independent of the position along the chain. A more general systematic derivation of the energy flux operator, based on conservation laws and valid for all phases of matter, is given in Refs. 31,35. Our present approach is different and, for example, makes it possible to study the energy resolved flux. The equivalence between our approach and that based on Eq. (38) is shown in Appendix C.

For definiteness we take the driving mode 0 to belong to the bath L. At steady state the energy flux into (and out of) the mode $r$ of the bath R is given by the quantum analog of (22), i.e.

$$J_{0 \to r} = \eta\omega_r \langle a_r^\dagger a_r + a_r a_r^\dagger - a_r a_r - a_r^\dagger a_r^\dagger \rangle_t / 2 \qquad (39)$$

is the flux transmitted through mode $r$,[36] where again $\langle \ \rangle_t$ denotes time average (in our application we also require averaging over the initial distribution of the driving mode states). Note that all operators here and below are Heisenberg representation operators at time $t$. Using Eq. (28) we obtain[29]

$$\frac{d}{dt}\left(a_r^\dagger a_r\right) = -i\sum_k \frac{V_{r,k}}{2\sqrt{\omega_k \omega_r}}\left[\left(a_k^\dagger + a_k\right)a_r - a_r^\dagger\left(a_k^\dagger + a_k\right)\right] - \eta/2\left[\left(a_r^\dagger - a_r\right)a_r + a_r^\dagger\left(a_r - a_r^\dagger\right)\right]$$
(40a)

$$\frac{d}{dt}\left(a_r a_r^\dagger\right) = -i\sum_k \frac{V_{r,k}}{2\sqrt{\omega_k \omega_r}}\left[a_r\left(a_k^\dagger + a_k\right) - \left(a_k^\dagger + a_k\right)a_r^\dagger\right] - \eta/2\left[a_r\left(a_r^\dagger - a_r\right) + \left(a_r - a_r^\dagger\right)a_r^\dagger\right]$$
(40b)

At steady state, the time average of Eq. (40) vanishes. This yields, using (39)

$$J_{0\to r} = i\omega_r \sum_k \frac{V_{r,k}}{4\sqrt{\omega_k \omega_r}}\left\langle \left(a_k^\dagger + a_k\right)\left(a_r^\dagger - a_r\right) + \left(a_r^\dagger - a_r\right)\left(a_k^\dagger + a_k\right)\right\rangle_t$$
(41)

Note that the dependence on the driving mode 0, while not explicit in (41), enters through the forms of the Heisenberg operators $a_k$ and $a_r$ that are solutions to Eqs. (33) and (37). The energy flux carried by modes in the range $\omega_0 ... \omega_0 + d\omega_0$ is given by $J_{L\to r}(\omega_0)d\omega_0$ where

$$J_{L\to r}(\omega_0) = i\omega_r \rho_L(\omega_0)\sum_k \frac{V_{r,k}}{4\sqrt{\omega_k \omega_r}}\left\langle \left(a_k^\dagger + a_k\right)\left(a_r^\dagger - a_r\right) + \left(a_r^\dagger - a_r\right)\left(a_k^\dagger + a_k\right)\right\rangle$$
(42)

where $\rho_L(\omega_0)$ is the density of modes of the left heat bath at frequency $\omega_0$. Using Eqs. (29)-(31) this leads to

$$J_{L\to r}(\omega_0) = \omega_0 \rho_L(\omega_0)\,\text{Im}\sum_k \frac{V_{r,k}}{2\sqrt{\omega_k \omega_r}}\left[<B_r A_k> + <A_k B_r>\right]$$
(43)

where terms such as $\langle A_k A_r\rangle e^{2i\omega_0 t}$ or $\langle B_k B_r\rangle e^{-2i\omega_0 t}$ that will yield zero average flux were disregarded. Next, using (37) to express $A_r$ and $B_r$ in terms of the $\{A_k\}$ and $\{B_k\} = \{A_k^\dagger\}$ operators, and taking the damping term $\eta$ there to zero, Eq. (43) leads to

$$J_{L\to r}(\omega_0) = \frac{\pi}{2}\rho_L(\omega_0)\delta(\omega_r - \omega_0)\sum_{k,k'} \frac{V_{r,k}V_{r,k'}}{2\sqrt{\omega_k \omega_{k'}}}\left[\left\langle A_k(\omega_0)A_{k'}^\dagger(\omega_0)\right\rangle + \left\langle A_{k'}^\dagger(\omega_0)A_k(\omega_0)\right\rangle\right]$$
(44)

To obtain (44) we have used the fact that $\sum_{k,k'}(\omega_k \omega_{k'})^{-1/2} V_{r,k} V_{r,k'} <A_k B_{k'}>$ is real and have disregarded a term that contains $\delta(\omega_r + \omega_0)$. We have also noted explicitly the fact that the $\{A_k\}$ operators, obtained from (33), depend on the driving frequency.

Eq. (44) shows, as expected in a linear system, that a driving (incoming) mode at frequency $\omega_0$ can excite outgoing modes only at this same frequency. The overall current per unit frequency range, transmitted from L to R at frequency $\omega_0$, is obtained by summing (44) over all final levels $\{r\}$:

$$J_{L \to R}(\omega_0) = \sum_r J_{L \to r}(\omega_0) =$$
$$= \frac{\pi}{2} \rho_L(\omega_0) \rho_R(\omega_0) \sum_{k,k'} \frac{(V_{r,k} V_{r,k'})_{\omega_r = \omega_0}}{2\sqrt{\omega_k \omega_{k'}}} \left[ \langle A_k(\omega_0) A_{k'}^\dagger(\omega_0) \rangle + \langle A_{k'}^\dagger(\omega_0) A_k(\omega_0) \rangle \right] \quad (45)$$

Note that the only attribute of this expression that makes it a "left-to-right" current is our initial designation of the driving mode as belonging to the left heat reservoir. The expectation values in (45) therefore depend on the temperature $T_L$ of the left bath. A similar expression with $T_R$ replacing $T_L$ applies for the right-to-left heat current.

From Eq. (33) it follows that one can write

$$A_k(\omega_0) = \bar{A}_k(\omega_0) V_{0,k} a_0^\dagger \sqrt{\frac{\omega_k}{\omega_0}} \quad (46)$$

where $\bar{A}_k(\omega_0)$ is a scalar function of the driving frequency that depends only on molecular parameters. The total heat current is obtained as the integral over all frequencies of the net current $J \equiv J_{R \to L} - J_{L \to R}$. Denote

$$\mathcal{T}(\omega) = \frac{\pi}{2} \rho_L(\omega) \rho_R(\omega) \sum_{k,k'} \frac{(V_{r,k} V_{r,k'})_{\omega_r = \omega} (V_{l,k} V_{l,k'})_{\omega_l = \omega}}{\omega^2} \left( \bar{A}_k(\omega) \bar{A}_{k'}^\dagger(\omega) + \bar{A}_{k'}^\dagger(\omega) \bar{A}_k(\omega) \right)/2$$

(47)

Using the definition of the friction from (35), we get

$$\mathcal{T}(\omega) = \frac{2\omega^2}{\pi} \sum_{k,k'} \gamma_{k,k'}^R(\omega) \gamma_{k,k'}^L(\omega) \left( \bar{A}_k(\omega) \bar{A}_{k'}^\dagger(\omega) + \bar{A}_{k'}^\dagger(\omega) \bar{A}_k(\omega) \right)/2 \quad (48)$$

The directional heat currents are therefore

$$J_{L \to R} = \int \mathcal{T}(\omega)(n_L(\omega) + 1/2) \omega d\omega \quad ; J_{R \to L} = \int \mathcal{T}(\omega)(n_R(\omega) + 1/2) \omega d\omega \quad (49)$$

and the net heat flux is

$$J = \int \mathcal{T}(\omega)[n_R(\omega) - n_L(\omega)] \omega d\omega \quad (50)$$

Which is our final result, similar to results recently derived in Refs. 4 and 30. The advantage of the present derivation (to be explored elsewhere) is that it can be easily generalized to any kind of initial boson distribution in the two baths, including driving by an external photon field. Expression (50) is similar to the Landauer result $J_{el} = \int \mathcal{T}(E)[f_R(E) - f_L(E)] dE$ for the electrical current in a junction connecting two electron reservoirs characterized by Fermi distributions $f_R(E)$ and $f_L(E)$ and a transmission function $\mathcal{T}(E)$.

We conclude this Section with two remarks. First, as already noted, the same result as given by Eqs. (45)-(50) can be obtained from the more conventional approach based on Eq. (38) (see Appendix C). Secondly, even though our treatment was described in the framework of a 1-dimensional molecule-bath coupling, the results are valid for a 3-dimensional coupling model: the needed input are the coupling elements between all the molecular normal-modes {$k$} and the phonons of the thermal baths {$l$} and {$r$}. See Appendix A for details.

## 4. Results and discussion

We next apply the formalism described above to the calculation of phonon induced heat transfer thermal conductance of a molecular bridge connecting two identical thermal reservoirs at different temperatures. We study alkane chains of variable length, and compare their heat transport properties to other ordered and disordered chains. The information needed for any given molecular bridge is the normal mode spectrum of the molecular system and the corresponding transformation matrix **C** (cf. Eq. (7)). These were obtained using the Hyperchem 6 computer package, with the isolated molecular geometry optimized using the Restricted Hartree-Fock method with the semi-empirical PM3 parameterization method. The index $N$ that denotes the molecule length is the number of backbone atoms, i.e. the carbones for the alkane systems. The parameters that characterize the reservoirs are the Debye cutoff frequency $\omega_c$, which is taken in the range $\omega_c$ =200-800 cm$^{-1}$, and the temperatures $T_L$ and $T_R = T_L + \Delta T$ which are studied in the range 10-1000K. Unless otherwise stated, $\Delta T$ itself is taken small, typically $\Delta T$= 10$^{-3}$K, so $T$ represents the average temperature of the two reservoirs.

Next consider the molecule-reservoirs coupling. We assume, as in Eq. (5), that it is affected by the extreme end-atoms on the two molecular edges. This coupling is commonly characterized by the spectral density function, e.g. between atom 1 and the left reservoir

$$d_L(\omega) = \frac{\pi}{2}\sum_l \frac{g_{1,l}^2}{m_l \omega_l}\delta(\omega-\omega_l) = \frac{\pi}{2}\frac{g_L^2(\omega)\rho_L(\omega)}{\omega m_L(\omega)} \tag{51}$$

where $\rho_L(\omega)$ is the mode density. The spectral density $d_L(\omega)$ is related to the frequency dependent friction on atom 1, Eq. (34), by

$$\gamma_L(\omega) = \frac{d_L(\omega)}{m_1 \omega} \tag{52}$$

where $\gamma_L(\omega)$ was defined by Eq. (36). In what follows we will assume that the spectral properties and coupling strengths are the same on left and right contacts and omit the indices $L$ and $R$ from $g(\omega)$, $\gamma(\omega)$, $\rho(\omega)$ and $m(\omega)$. We use a Debye-like model defined by

$$\rho(\omega) = N_B \frac{\omega^2}{2\omega_c^3} e^{-\omega/\omega_c} \tag{53}$$

where $N_B$ is the number of reservoir modes. This leads to

$$d(\omega) = a m_1 \omega e^{-\omega/\omega_c}; \quad \gamma(\omega) = a e^{-\omega/\omega_c} \tag{54}$$

where (from (53) and (51))

$$a = \frac{\pi g^2(\omega) N_B}{4 m_1 m(\omega) \omega_c^3} \tag{55}$$

Here $m_1 = m_N$ is the mass of the end atom on the molecular chain. Further simplification is achieved by considering atomic baths for which $m(\omega) = m_B$ and by assuming that $g(\omega) = g$ does not depend on $\omega$. The magnitude of g measures the strength of the molecule-bath coupling. In a model where we take the coupling between the molecular chain and the thermal reservoirs to be dominated (or gated) by the coupling between the end chain atoms (1 and $N$) and their nearest neighbor atoms ($L$ and $R$, say) in the corresponding reservoirs, we may write this coupling (in correspondence with Eqs. (3) and (5)) as (focusing for definiteness of notation on the left reservoir)

$$H_{ML} = -g_{1,L} x_1 x_L = -\sum_l g_{1,l} x_1 x_l \tag{56}$$

The second equality results from expanding the local coordinate $x_L$ of the reservoir atom in the reservoir normal modes, $x_L = \sum_l \alpha_l x_l$ ($\sum_l |\alpha_l|^2 = 1$). This implies

$$g_{1,l} = g_{1,L}\alpha_l \tag{57}$$

so that $\sum_l g_{1,l}^2 = g_{1,L}^2$, or, if $g_{1,l} = g$,

$$g = \frac{g_{1,L}}{\sqrt{N_B}} \tag{58}$$

With these simplifications Eq. (55) takes the form

$$a = \frac{\pi}{4} \frac{g_{1,L}^2}{m_1 m_L \omega_c^3} \tag{59}$$

In the calculations described below this constant is taken in the range $10^4$-$10^5$ cm$^{-1}$.

Once the normal mode spectrum and the transformation matrix **C** (Eq. (7)) have been calculated, Eq. (10), (33), (35), (47) and (50) are used to calculate the heat flux and the heat conductance. The latter is defined by

$$K = \lim_{\Delta T \to 0} J / \Delta T \tag{60}$$

The thermal conductivity of 1-dimensional atomic chains and its dependence on the chain length was studied before by several groups.[5,6,30] It was found[5,6] that in a perfect harmonic chain the heat flux is not proportional to the temperature gradient $(T_R - T_L)/N$, as inferred from Fourier law, but to the temperature difference $T_R$-$T_L$. The thermal conductance, $J/\Delta T$, was predicted to be independent of the chain length, and the thermal conductivity for unit cross-sectional area, $J/\nabla T$, therefore diverges as the chain length goes to infinity. Motion in our molecular chains is not restricted to one dimension, still the proximity of these chains to the 1-dimensional models suggests perhaps a similar behavior.

Figure 2 shows the dependence of the calculated heat conductance on chain length for linear alkanes of 2-25 carbon atoms at different temperatures. The molecule-reservoir coupling parameter was taken $a$=8000 cm$^{-1}$. The conductance becomes length independent for $N>15$, while for short chains, $N$=2-4, we see an unexpected rise of the conductance with chain length. The inset shows a similar result for the strong coupling case, $a$=1.2·10$^5$ cm$^{-1}$, and $T$=1000K. Here the heat conductance appears to decrease like $K \propto 1/N$ for large $N$.

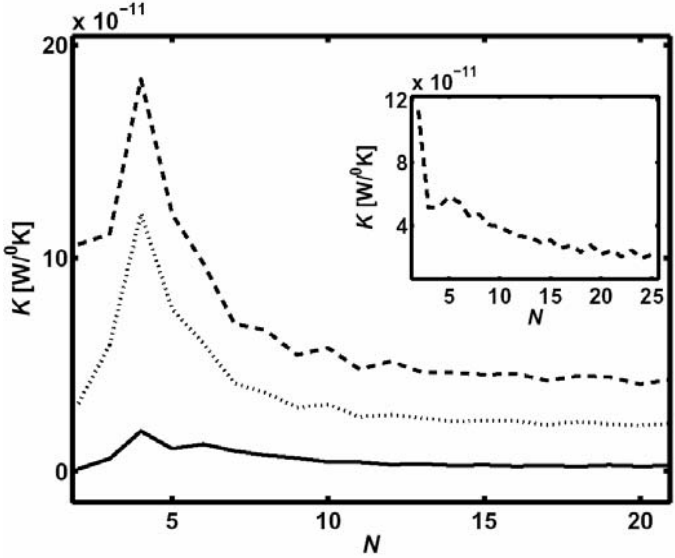

FIG 2: The thermal conductance calculated as a function of chain length for alkane molecules at different temperatures, using $\omega_c$=400 cm$^{-1}$ and $a$=8000 cm$^{-1}$. Full line: $T$=50K; dotted line: $T$=300K; dashed line: $T$=1000K. The inset shows the $T$=1000K result in the case of strong molecule-reservoirs coupling, $a$=1.2·10$^5$ cm$^{-1}$ (and same $\omega_c$=400 cm$^{-1}$).

The asymptotic dependence of the heat conductance on the chain length is of particular interest. Fourier's law of heat conductance would imply a $1/N$ dependence on chain length, while simple arguments based on Eq. (47) suggest that the conductance should be length independent for long chains. The argument is that the phononic transmission $\mathcal{T}(\omega)$ depends on the coupling strength with a fourth power, contributing the factor $1/N^2$, while the double sum in (47) yields the factor $\rho_M(\omega)^2$, where $\rho_M(\omega)$, the molecular density of states, increases linearly with the molecule length. If the possibility that the factor $\left(\overline{A}_k(\omega)\overline{A}_{k'}^\dagger(\omega)\right)$ in (47) may depend on this length is disregarded, the heat flux $J$ is expected to be length independent. The actual answer to this issue depends on the density and the localization properties of the molecular normal modes.

Figures 3 and 4 display some properties of the normal modes in alkane chains. Figure 3 depicts the density of modes for chains with $N$=15 and $N$=30 atoms. Three domains, separated by gaps, are seen in the spectrum. A group of low frequency modes in the range below 600 cm$^{-1}$, intermediate frequency modes with $\omega$=700-1500 cm$^{-1}$ and high frequency modes of $\omega$=2950-3200 cm$^{-1}$. Note that the modes in the intermediate region have the highest density of states, and that the modes density increases linearly with the molecular size.

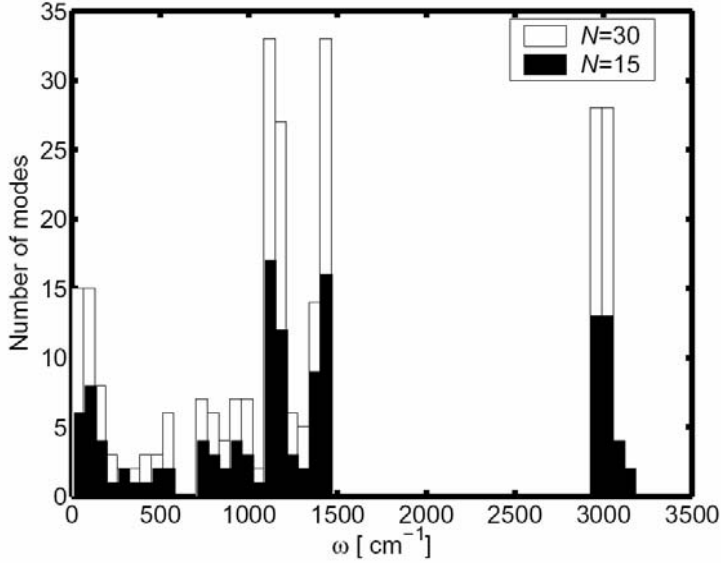

FIG 3: The spectral distribution of normal modes of alkane chains of lengths $N=15$ and $N=30$.

In order to gain a better understanding of these normal modes we follow previous work on heat transfer in disordered harmonic media.[35,37]. In particular, the ability of any mode to transfer energy across the molecule depends on its spatial extent, which may be characterized by the participation ratio $P_k$. In the present application we use a variation of the measure suggested by[38]. Define the weight associated with normal mode $k$ on the carbon segment $n$ as the sum

$$p_{k,n} = \sum_{\alpha_n} |(C^{-1})_{k,\alpha_n}|^2 \qquad (61)$$

where $\alpha_n$ goes over all atoms (hydrogens and carbon) associated with a given carbon atom. Note that $\sum_n p_{k,n} = 1$. The participation ratio is given by

$$P_k = \left[ \sum_n p_{k,n}^2 \right]^{-1} \qquad (62)$$

This is a good measure for the number of carbon sites on which the collective mode $k$ has a significant amplitude. For a chain of $N$ carbon atoms, $P_k = N$ for a ballistic mode that extends over the entire molecule, and it decreases as localization becomes more significant. An alternative measure is the information entropy[39]

$$S_k = -\sum_{n=1}^{N} p_{k,n} \ln(p_{k,n}) \tag{63}$$

which satisfies $S_k = \ln(N)$ for a completely delocalized mode, and $S_k = 0$ for a mode localized on a single site.

Figure 4 depicts the average participation ratio $\langle P \rangle$ and the average function $\langle \exp(S) \rangle$ for each group of modes plotted against the chain length. Both measures increase linearly with chain length in all cases, indicating some ballistic nature for at least some modes in each group, yet the high frequency modes are, on the average, more localized. In contrast, the low frequency modes show a substantial delocalized character.

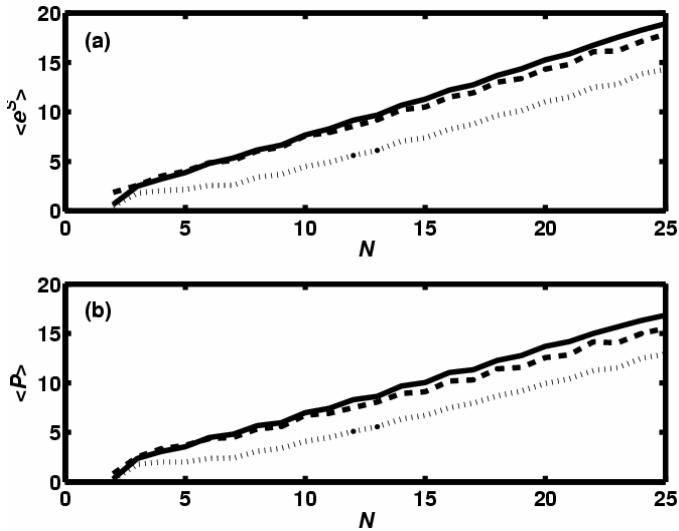

FIG. 4: Measures of mode localization in alkanes plotted as functions of molecular length. (a) The function $e^S$ where $S$ is the information entropy, Eq. (63), averaged over a group of modes as indicated below. (b) The average participation ratio, Eq. (62), for the same groups. Dashed line: low frequency modes ($\omega$<600cm$^{-1}$); full line: intermediate frequency modes, $700 \leq \omega \leq 1500$ cm$^{-1}$; dotted line: high frequency modes, $\omega = 2950 \leq \omega \leq 3200$ cm$^{-1}$

The interplay between the number of modes in each frequency group, their ability to transfer energy as derived from their localization property, and the frequency dependence of the mode population in the thermal reservoirs combine to affect the chain-length dependence of heat transport in our model junction. This can be seen by studying separately the heat conduction behavior of the three frequency groups. It should be emphasized that the contributions of different modes to the heat conduction is not additive, as can be seen from the presence of cross terms in Eq. (47). Still, looking at these separate contributions provides useful insight, and in fact describing the overall heat conduction as an additive combination of contributions from the three frequency groups defined above is found (see below) to be a good approximation for long enough chains.

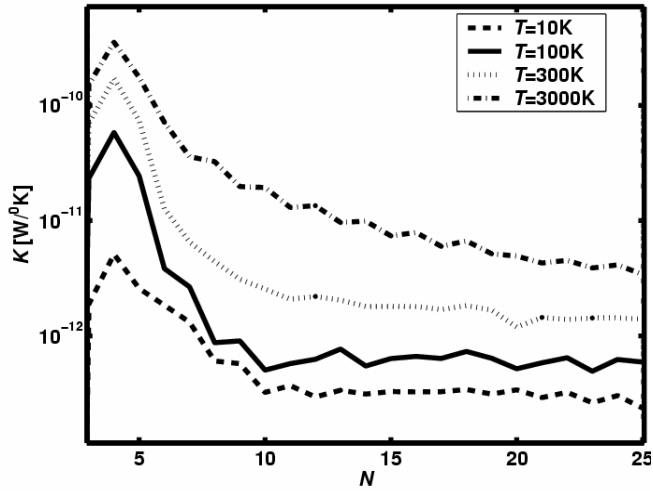

FIG. 5: Thermal conductance vs. chain length for alkane chains in which only the low-frequency modes ($\omega<600\,\mathrm{cm}^{-1}$) are taken into account. $\omega_c=400\,\mathrm{cm}^{-1}$, $a=1.2\cdot10^5\,\mathrm{cm}^{-1}$. Dashed line: $T=10\mathrm{K}$; Full line: $T=100\mathrm{K}$; dotted line: $T=300\mathrm{K}$; dash-dotted line $T=3000\mathrm{K}$.

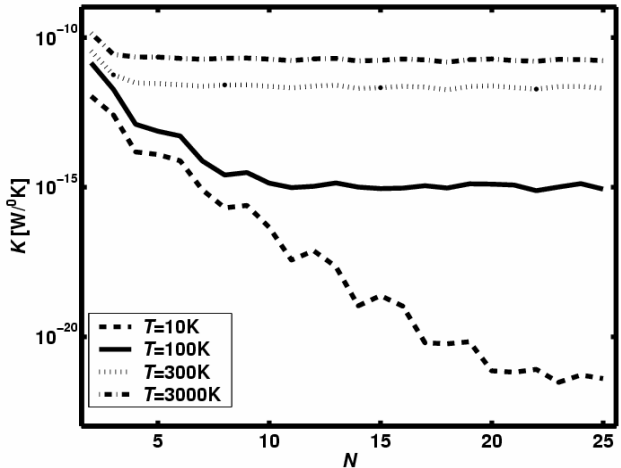

FIG. 6: Same as Fig. 5, except that only the intermediate-frequency modes ($700\leq\omega\leq 1500\,\mathrm{cm}^{-1}$) are taken into account.

Figures 5-6 show the heat conductance vs. alkane chain-length for the low and intermediate frequency modes respectively, at different temperatures, using $\omega_c=400$cm$^{-1}$ and $a=1.2\cdot10^5$ cm$^{-1}$ (same parameters as in the inset of Fig. 2). The following observations can be made:

(1) The conductance associated with the low frequency modes (Fig. 5) does not depend on the chain length at low temperatures (*T*), however it decreases with chain length in the high temperature regime.

(2) The intermediate frequency modes (Fig. 6) show a very different behavior: For low temperature the conductance decreases exponentially, while for high temperature, it becomes length independent.

(3) The high frequency modes (figure not shown) basically show the same behavior as that of the intermediate frequency group, with stronger variations about the systematic trend. However, the absolute contribution of these modes to the heat transfer is negligible as compared to the other two groups.

These different modes of behavior can be understood as transitions between two modes of transport: tunneling-like and resonant. Consider, for example, Fig. 6. At very low temperatures modes of the thermal reservoirs that are in resonance with the intermediate molecular modes considered here are not populated. Only low-frequency modes of the reservoirs are excited, and the transmission of the energy between these reservoirs through the molecule is a transfer of low frequency phonons through a bridge of relatively high frequency vibrations. This leads to a tunneling type behavior with an exponential decrease of the transmission with bridge length, in analogy with the super-exchange mechanism of electron transfer.[40] When the temperature increases, higher frequency modes of the reservoirs, which are in resonance with the intermediate molecular modes are excited and contribute to resonance transmission which is distance independent. Similar considerations apply in principle to the high frequency modes, but the contribution of these modes to the heat transfer is small because the Debye cutoff $\omega_c$ of the reservoir spectra is considerably below these modes.

Consider now Figure 5, which shows the chain-length dependence of the heat conduction by the low frequency molecular mode. Here we see an opposite behavior, where the heat conductance does not depend on chain-length at low *T*, while it decreases with chain length at high *T*. The same arguments as before apply also in this case. At low temperatures heat transfer involves low frequency reservoir modes that are in resonance with the molecular frequencies of this group, hence the length independence of the transmission. For high *T*, the high-frequency reservoir modes are activated, however transmission involving these modes is a non-resonance process that decreases with chain length. The dependence on length in this case is weaker than exponential because the thermal shift of population from low to high frequency reservoir modes is very gradual.

We emphasize these observations by comparing the heat conductance from the low and intermediate frequency molecular modes, calculated at the unphysical temperature *T*=3000K where their relative contributions are comparable, see the dashed-dotted lines in Figs 5-6. (The contribution of the high frequency molecular modes is small even at this unphysically high temperature). In this temperature range the conductance due to the low-frequency modes (figure 5) decreases with chain-length, that of the intermediate modes (figure 6) is length independent and their superposition is therefore expected to show a relatively weak length dependence. We have verified numerically that taking a simple superposition of these two contributions is indeed a good approximation to the full calculation for chains longer than 6-7 carbon atoms, and therefore this analysis holds. For very long chains the non-resonant contributions die out and heat transmission becomes length independent, though because only a few modes may be extended enough it may be small. The turnover to length dependence at smaller chain-lengths and the actual length dependence of the overall conduction for relatively short chains depends on the molecule-reservoir coupling strength, on the temperature and on phonon spectra of both molecule and reservoirs.

The dependence of the heat conductance on the temperature is depicted in Fig. 7. The system parameters used here are $a$=8000 cm$^{-1}$ and $\omega_c$=800 cm$^{-2}$. At the high temperature limit, shown in the inset, the conductance saturates to the value $k_B \int \mathcal{T}(\omega) d\omega$, that corresponds to the high temperature limit of equation (50). In this limit the conductance decreases with the chain length *N*. At the very low temperatures, main graph, the trend is reversed: the conductance increases with chain lengths for short chains.

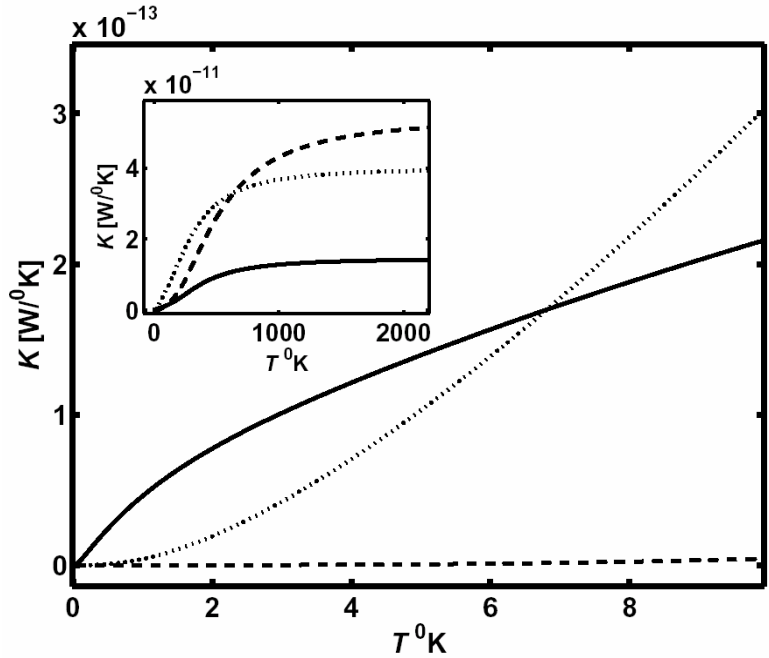

FIG.7: Thermal conductance calculated as a function of temperature for alkane chains using $\omega_c$=800 cm$^{-1}$ and $a$=8000 cm$^{-1}$. Dashed line: $N$=2; Dotted line: $N$=5; Full line: $N$=14. Inset shows the high temperature regime.

The conductance increase with longer chain length, seen in Fig. 7 and also on the short chain (left) sides of Figs. 2 and 5, seems at first counter intuitive, however Figure 8 reveals its origin. Here we show the thermal conductance as a function of chain length at several temperatures. Variation of the chain length affects the molecular normal mode spectrum in two ways. First, the overall density of states is increased linearly. Secondly, the lower bound on this density is shifted to lower values. For example, for a pentane ($N$=5) the lowest vibrational frequency mode is $\omega$=84 cm$^{-1}$, for decane ($N$= 10) it is $\omega$=28 $cm^{-1}$, while for $N$=20 it is $\omega$ =7 $cm^{-1}$. At low temperatures the heat current is carried mostly by low frequency phonons, and when the chain becomes longer, more molecular modes come into resonance with these incoming phonons. This causes an increase in the heat flux. Obviously, this effect should be significant only at very low temperatures, as indeed seen in Fig. 8.

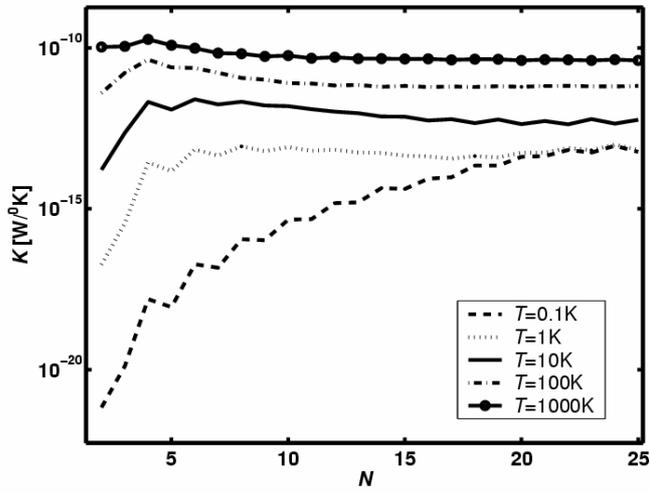

FIG. 8: Thermal conductance calculated as a function of length for alkane chains, using $\omega_c$=400 cm$^{-1}$ and $a$=8000 cm$^{-1}$. Dashed line: $T$=0.1K; dotted line: $T$=1K; full line: $T$=10K; dash-dotted line: $T$=100K; line with filled circles: $T$=1000K.

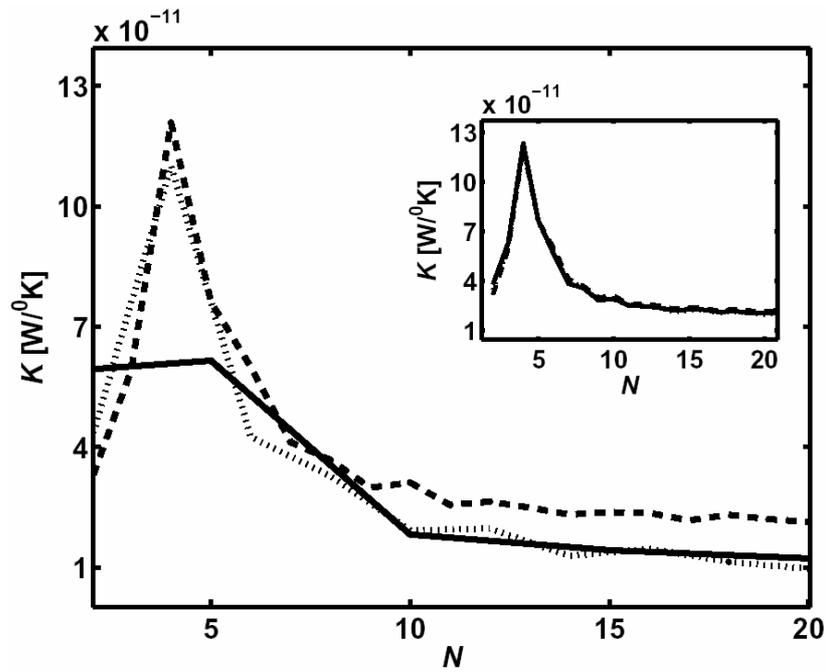

FIG. 9: Thermal conductance vs. alkane chain length. $\omega_c$=400 cm$^{-1}$, T=300K, $a$=8000 cm$^{-1}$. Dashed line: $^{12}$C chains; full line: $^{28}$C chains; dotted line: disordered $^{12}$C – $^{28}$C chains with 1:1 component ratio. The inset depicts similar results for the case where $^{14}$C replaces $^{28}$C.

*Disordered chains.* Figure 9 compares the heat conduction of pure alkane chains to similar chains with a random distribution of two masses with 1:1 component ratio. $\omega_c$=400 cm$^{-1}$, $a$=8000 cm$^{-1}$, and T=300K were used in these calculations. The chains are normal alkanes in which the atomic masses of half the carbon atoms have been set artificially to 28. $^{12}$C/$^{28}$C case. We see that for a long enough chain, the heavy atom chains with smaller normal mode frequencies conduct less effectively than their light atom analogs. This results from the balance of three effects. First, the contributions of modes of different frequencies depend on the corresponding reservoirs density of mode spectra. Secondly, it depends on the thermal populations of these modes. Finally, the energy carried by a mode of frequency $\omega$ is proportional to $\omega$. The effect of disorder also involves balancing factors: Starting from, e.g., the $^{28}$C chain and replacing some of these heavy atoms with the $^{12}$C isotope would reduce heat conduction because of localization (for example, in a 20 carbon chain the localization measure $\langle \exp S \rangle$, Eq. (63) averaged over all modes, is 14.1 for a pure $^{12}$C chain, 12.5 for $^{28}$C system, and only 7.5 for the random $^{12}$C/$^{28}$C chain). This is partly balanced by the shift of the mixed structures spectrum into frequencies above those of the pure $^{28}$C chain. In Figure 9 this results in little difference in the heat conduction of a pure $^{28}$C chain and a $^{12}$C/$^{28}$C mixed chain.

Finally we note that a similar behavior is seen for the realistic $^{12}$C/$^{14}$C chains, however the difference between the heat conductions of the pure and the mixed chains in this case is quite small and are hardly resolved on the scale of Fig. 9 (see inset).

*Anharmonic effects.* In macroscopic systems and in fact whenever the system size is larger than the localization length and/or mean free path (determined by disorder and scattering by anharmonic interactions) heat conduction is dominated by anharmonic coupling. In our short molecular chains such effects are expected to play a much lesser role, at least at low temperature. To examine this issue we have carried out classical numerical simulations of heat conduction through a 1-dimensional model of alkane chains without invoking the harmonic approximation. Details of the model and the calculation are provided in Appendix D.

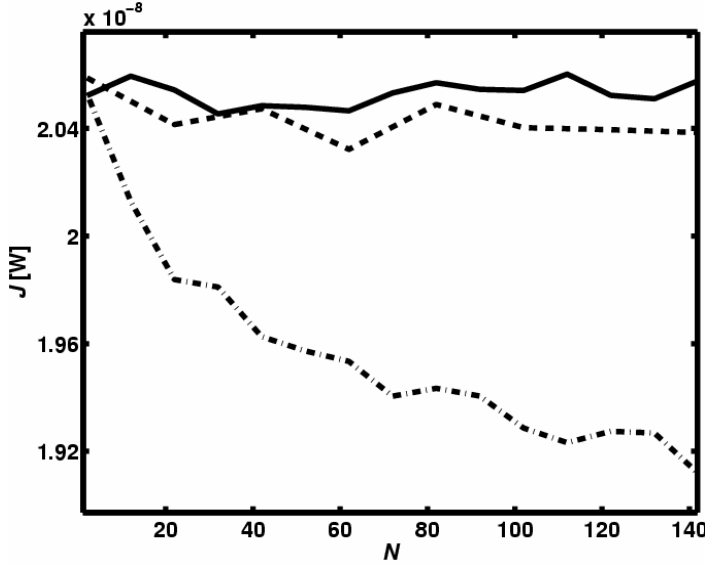

FIG. 10: Heat current vs. chain length obtained from a classical simulation of heat transport through 1-dimensional model alkane chains characterized by different anharmonic interactions. Full line: harmonic chain; dashed line: anharmonic chain using the alkane force field parameters; dash-dotted line: anharmonic chain with unphysically large anharmonicity ($\xi$ of Eq. (D2) taken 36 times the alkane value). $\gamma=10\text{ps}^{-1}$ ($\gamma$ is the friction coefficient defined in Appendix D), $T_R$=300K, $T_L$=0K were used in this simulation.

Figure 10 shows the heat current obtained from such a calculation. For a long harmonic chain the heat flux is ballistic and does not depend on chain length. The result for the full alkane model (dashed line) and the corresponding harmonic approximation (full line) are seen to behave in this way and to be very close to each other. Only when the molecular anharmonicity is taken unphysically large (dashed-dotted line) we see deviation from this behavior and a decrease of the current with chain length. Similar deviations from the harmonic behavior (not shown) are seen at elevated temperatures, but only when $T$ is unphysically high, say $T \geq 2000$K.

*Comparison to classical heat transfer.* Finally, we compare the heat conduction properties of the harmonic chains considered in this paper with the continuum heat transport model used in Ref. 22. In that paper the molecule was represented by a cylinder connecting the two heat reservoirs (Fig. 1) and a continuum model was employed to estimate the heat conduction, using for the thermal conductivity coefficient the value $\sigma_h = 3.5 \cdot 10^{-4}$ cal/(s·cm·K) typical of bulk organic solids. For a model in which the molecular cylinder of length 60Å is suspended in vacuum between the two heat reservoirs at 300K a modest temperature rise of a few degrees was found when heat was deposited uniformly on the cylinder at a rate of $10^{10}$eV/s (corresponding to about 1nA electron current flowing across a potential bias of 1V). Clearly however, macroscopic heat conduction, dominated by impurity scattering and anharmonic interactions cannot reliably represent heat conduction of molecular junctions that is characterized by harmonic (ballistic) transport on one hand, and by restricted geometry and the availability of conducting modes on the other.

For definiteness we assume that the coupling between the molecular chain and the thermal reservoirs is dominated (or gated) by the coupling between two nearest neighbor alkane carbon atoms. This is expected to overestimate the actual thermal coupling in most molecular junctions. This implies that in Eq. (59) we take $g_{1,L} = g_{CC} \equiv g$ (similarly we take $g_{N,R} = g$). Also for definiteness we assign carbon masses to the baths, i.e. take $m_L = m_R = m_C$. The corresponding numerical values are $g = 7.2 \cdot 10^{-3}$ dyne/Å (from the Hyperchem force field) and $m_C = 2 \cdot 10^{-26}$ kg. This implies (from Eq. (59)) $a = 2 \cdot 10^7$ cm$^{-1}$ for $\omega_c = 400$ cm$^{-1}$, $a = 2.5 \cdot 10^6$ cm$^{-1}$ for $\omega_c = 800$ cm$^{-1}$ and $a = 5 \cdot 10^5$ cm$^{-1}$ for $\omega_c = 1400$ cm$^{-1}$. The latter $\omega_c$ is the order of the Debye frequency of diamond.

Numerical results obtained from this model are compared to the classical cylinder model of Ref. 22 are shown in Fig. 11. The classical calculation was done using $\tilde{K} = 3.5 \ 10^{-4} cal/(\sec \ cm \ ^0K)$ (=1.5 $10^{-11}$ W/(Å $^{\circ}$K)) for the heat conductivity coefficient, and a cross-sectional area $\mathcal{A} = 3.5$ Å$^2$ for the "molecular" cylinder. The length of the cylinder that corresponds to an alkane chain with $N$ carbon atoms was taken $L = 1.2N$ Å. Note that the heat conductance of this classical object, $K = \tilde{K}\mathcal{A}/L$ decreases as $N^{-1}$ with chain length. For $N$=5-20 we get that $K$=10$^{-11}$-10$^{-12}$ W/$^{\circ}$K.

The results displayed in Fig. 11 show that the heat conduction of the macroscopic cylinder overestimates that of the molecular model by about an order of magnitude at room temperature, while they are very similar at $T$=1000K. These observations are not very sensitive to details of the chosen coupling and reservoir cutoff parameters within a reasonable range. In view of the different mechanisms involved, one should not take the similar transport properties at 1000K as an approach to the classical limit at high *T*. More significant is the finding that at room temperature the classical model strongly overestimates the heat conduction properties of the individual molecule, an observation of important potential consequences for estimating heating in conjunction with electrical conduction in molecular junctions.

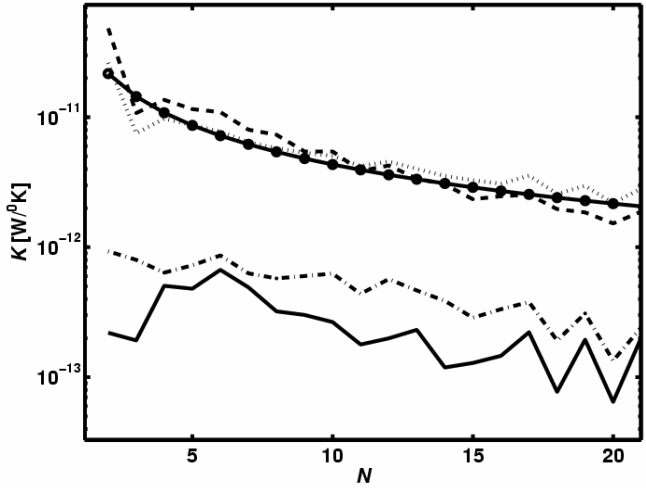

FIG. 11: Heat conduction vs. alkane chain length. Full line: $\omega_c$=400 cm$^{-1}$, $a$=2·10$^7$ cm$^{-1}$, $T$=300K. Dashed line: $\omega_c$=400 cm$^{-1}$, $a$=2·10$^7$ cm$^{-1}$, $T$=1000K. Dashed-dotted line: $\omega_c$=800 cm$^{-1}$, $a$=2.5·10$^6$ cm$^{-1}$, $T$=300K. Dotted line: $\omega_c$=800 cm$^{-1}$, $a$=2.5·10$^6$ cm$^{-1}$, $T$=1000K. Line with circles: Results of the classical continuum calculation (see text).

## 5. Conclusions

The heat conduction properties of molecular chains connecting two thermal reservoirs were investigated theoretically and numerically, focusing on saturated alkane chains as a primary example. It was found that heat conduction in relatively short chain is dominated by harmonic interactions. The harmonic approximation utilized yields a Landauer type expression (Eq. (50)) for the heat current, where energy is carried ballistically through the wire. The principal factors that determine heat conduction in such molecular junctions are the molecular vibrational spectral density, the localization properties of molecular normal modes in the different spectral regimes, the coupling of the molecule to the reservoirs and the cutoff frequency that characterizes the reservoirs spectral densities. The dependence of the heat conduction on molecular length varies with temperature and reflects the different localization properties of different molecular spectral regimes. Mode localization also causes disordered chains to be less effective heat conductors. A classical heat conduction model was found to overestimate the microscopic result by about one order of magnitude, an observation of potential importance for the estimate of heating associated with electrical conduction in molecular junction.

**Appendix A: The 3-dimesional case**

Generalization of the formulation presented in sections 2-3 into three dimensions is trivial. Since both molecule and reservoirs are described in terms of their normal modes, the dimensionality enters explicitly only in the form of the molecule-reservoir coupling. In three dimensions the Hamiltonian, Eq. (3), takes the form

$$H_B + H_{MB} = \sum_l \left\{ \frac{1}{2} m_l \omega_l^2 \left( \mathbf{r}_l - \frac{g_{1,l} \mathbf{r}_1}{m_l \omega_l^2} \right)^2 + \frac{\mathbf{p}_l^2}{2m_l} \right\} + \sum_r \left\{ \frac{1}{2} m_r \omega_r^2 \left( \mathbf{r}_r - \frac{g_{N,r} \mathbf{r}_N}{m_r \omega_r^2} \right)^2 + \frac{\mathbf{p}_r^2}{2m_r} \right\}$$

(A1)

where $\mathbf{r}$ and $\mathbf{p}$ are three components vectors. For simplicity we take the coupling strength $g$ to be the same for the three directions. The transformation matrix $\mathbf{C}$ (Eq.(7)) in this case is a *3N* x *3N* matrix, $\mathbf{r} = \mathbf{C}\bar{\mathbf{r}}$, where any consecutive three components of the vector r represent the *x,y,* and *z* coordinates of an atom in the molecule. The coupling terms are defined similarly to Eq. (10) as

$$V_{l,k}^x \equiv \frac{g_{1,l} C_{1,k}}{\sqrt{m_1 m_l}} \; ; \; V_{l,k}^y \equiv \frac{g_{1,l} C_{2,k}}{\sqrt{m_1 m_l}} \; ; \; V_{l,k}^z \equiv \frac{g_{1,l} C_{3,k}}{\sqrt{m_1 m_l}}$$

$$V_{r,k}^x \equiv \frac{g_{N,r} C_{3N-2,k}}{\sqrt{m_N m_r}} \; ; \; V_{r,k}^y \equiv \frac{g_{N,r} C_{3N-1,k}}{\sqrt{m_N m_r}} \; ; \; V_{r,k}^z \equiv \frac{g_{N,r} C_{3N,k}}{\sqrt{m_N m_r}}$$

(A2)

The real parts of the damping terms, Eq. (34), are given by

$$\gamma_{k,k'}^L(\omega) = \frac{\pi}{2} \sum_l \frac{\left( V_{l,k}^x V_{l,k'}^x + V_{l,k}^y V_{l,k'}^y + V_{l,k}^z V_{l,k'}^z \right)}{\omega_l^2} \delta(\omega - \omega_l)$$

$$\gamma_{k,k'}^R(\omega) = \frac{\pi}{2} \sum_r \frac{\left( V_{r,k}^x V_{r,k'}^x + V_{r,k}^y V_{r,k'}^y + V_{r,k}^z V_{r,k'}^z \right)}{\omega_r^2} \delta(\omega - \omega_r)$$

(A3)

And the operators $A_k$ satisfy the equation

$$\left(\omega_k^2 - \omega_0^2 + i\left[\gamma_{k,k}^L(\omega_0) + \gamma_{k,k}^R(\omega_0)\right]\omega_0\right)A_k(\omega_0) +$$
$$+ i\sum_{k'\neq k}(\gamma_{k,k'}^L(\omega_0) + \gamma_{k,k'}^R(\omega_0))\omega_0\sqrt{\frac{\omega_k}{\omega_{k'}}}A_{k'}(\omega_0) = \left[V_{0,k}^x a_{0,x}^\dagger + V_{0,k}^y a_{0,y}^\dagger + V_{0,k}^z a_{0,z}^\dagger\right]\sqrt{\frac{\omega_k}{\omega_0}} \quad (A4)$$

where the average of the driving operators, the analog of Eq. (27), is given by

$$\left\langle a_{0,j}^\dagger a_{0,j'}\right\rangle_L = n_L(\omega_0)\delta_{j,j'} \quad ; \quad \left\langle a_{0,j}^\dagger a_{0,j'}\right\rangle_R = n_R(\omega_0)\delta_{j,j'} \qquad j,j'=\{x,y,z\} \quad (A5)$$

Finally the transmission is given by summing the coupling to the reservoirs in Eq.(47) over the three spatial dimensions,

$$T(\omega) = \frac{\pi}{2}\rho_L(\omega)\rho_R(\omega)\sum_{j,j'}\sum_{k,k'}\frac{\left(V_{r,k}^j V_{r,k'}^j\right)_{\omega_r=\omega}\left(V_{l,k}^{j'} V_{l,k'}^{j'}\right)_{\omega_l=\omega}}{\omega^2}\left(\overline{A}_k(\omega)\overline{A}_{k'}^\dagger(\omega) + \overline{A}_{k'}^\dagger(\omega)\overline{A}_k(\omega)\right)/2$$
(A6)

and the net heat current is given by the same expression as Eq. (50).

## Appendix B: The generalized Langevin equation

The derivation of the generalized Langevin equation (GLE) starts with equation (12). Laplace transforming with $t_0$ as the initial time yields

$$(s^2 + \omega_l^2)\overline{x}_l(s) = \dot{\overline{x}}_l(t_0) + s\overline{x}_l(t_0) + \sum_k V_{l,k}\overline{x}_k(s) \quad (B1)$$

where

$$\overline{x}(s) = \int_{t_0}^\infty e^{-s(t-t_0)}\overline{x}(t)dt \quad . \quad (B2)$$

Rearrangement of (B1) leads to

$$\overline{x}_l(s) = \frac{s\overline{x}_l(t_0)}{s^2 + \omega_l^2} + \frac{\dot{\overline{x}}_l(t_0)}{s^2 + \omega_l^2} + \sum_k \frac{V_{l,k}}{s^2 + \omega_l^2}\overline{x}_k(s) \quad (B3)$$

and transforming back into the time domain produces

$$\overline{x}_l = \overline{x}_l(t_0)\cos(\omega_l(t-t_0)) + \frac{\dot{\overline{x}}_l(t_0)\sin(\omega_l(t-t_0))}{\omega_l} + \sum_k \frac{V_{l,k}}{\omega_l}\int_{t_0}^t \overline{x}_k(\tau)\sin(\omega_l(t-\tau))d\tau \quad (B4)$$

The last equation was derived using the convolution relation

$$\int_{t=t_0}^\infty e^{-s(t-t_0)}dt\int_{\tau=t_0}^t g(\tau)f(t-\tau)d\tau = g(s)f(s). \quad (B5)$$

Integrating Eq. (B4) by parts leads to

$$\bar{x}_l = \bar{x}_l(t_0)\cos(\omega_l(t-t_0)) + \frac{\dot{\bar{x}}_l(t_0)\sin(\omega_l(t-t_0))}{\omega_l} + \sum_k \frac{V_{l,k}}{\omega_l^2}\bar{x}_k(t)$$
$$-\sum_k \frac{V_{l,k}}{\omega_l^2}\bar{x}_k(t_0)\cos(\omega_l(t-t_0)) - \sum_k \frac{V_{l,k}}{\omega_l^2}\int_{t_0}^{t}\ddot{\bar{x}}_k(\tau)\cos(\omega_l(t-\tau))d\tau \quad (B6)$$

Now insert (B6) into Eq. (11). For clarity we ignore for the moment the coupling to the right reservoir. This yields

$$\ddot{\bar{x}}_k = -\omega_k^2\bar{x}_k + \sum_l V_{l,k}\left[\bar{x}_l(t_0)\cos(\omega_l(t-t_0)) + \frac{\dot{\bar{x}}_l(t_0)\sin(\omega_l(t-t_0))}{\omega_l}\right] -$$
$$\sum_l V_{l,k}\left\{\sum_{k'}\frac{V_{l,k'}}{\omega_l^2}\bar{x}_{k'}(t_0)\cos(\omega_l(t-t_0)) + \sum_{k'}\frac{V_{l,k'}}{\omega_l^2}\int_{t_0}^{t}\ddot{\bar{x}}_{k'}(\tau)\cos(\omega_l(t-\tau))d\tau\right\} \quad (B7)$$

Because we will be interested in the steady state of a linear system affected by damping interactions the term involving $\bar{x}_{k'}(t_0)$ can be ignored, since the initial conditions of the system are not relevant, see also [41,42]. The terms containing $\bar{x}_l(t_0)$ and $\dot{\bar{x}}_l(t_0)$ are recognized a harmonic driving force

$$\tilde{x}_l(t) \equiv \bar{x}_l(t_0)\cos(\omega_l(t-t_0)) + \frac{\dot{\bar{x}}_l(t_0)\sin(\omega_l(t-t_0))}{\omega_l} \quad (B8)$$

that obeys the harmonic oscillator equation of motion $\ddot{\tilde{x}}_l = -\omega_l^2\tilde{x}_l$. Similar contributions are obtained for the right reservoir, yielding finally

$$\ddot{\bar{x}}_k = -\omega_k^2\bar{x}_k + \sum_l V_{l,k}\tilde{x}_l + \sum_r V_{r,k}\tilde{x}_r$$
$$-\sum_{l,k'}\frac{V_{l,k}V_{l,k'}}{\omega_l^2}\int_{t_0}^{t}\ddot{\bar{x}}_{k'}(\tau)\cos(\omega_l(t-\tau))d\tau - \sum_{r,k'}\frac{V_{r,k}V_{r,k'}}{\omega_r^2}\int_{t_0}^{t}\ddot{\bar{x}}_{k'}(\tau)\cos(\omega_r(t-\tau))d\tau \quad (B9)$$

where

$$\tilde{x}_r(t) \equiv \bar{x}_r(t_0)\cos(\omega_r(t-t_0)) + \frac{\dot{\bar{x}}_r(t_0)\sin(\omega_r(t-t_0))}{\omega_r} \quad (B10)$$

obeys $\ddot{\tilde{x}}_r = -\omega_r^2\tilde{x}_r$.

The procedure outlined above is a standard derivation of a generalized Langevin equation to be used in the long time limit of the system-bath interaction where initial system information can be neglected. A different approach can be used to get directly the steady-state equations for a driven quantum system. This formulation uses the ideas of sections 2-3: the quantization of the momentum and displacement, Eqs. (23)-(27), driven steady-state dynamics, Eq.(29) and working in the frequency domain. We focus on the left reservoir and start again from Eq. (12), but replaced by its damped analog, Eq. (21): $\ddot{\bar{x}}_l = -\omega_l^2 \bar{x}_l + \sum_k V_{l,k} \bar{x}_k - \eta \dot{\bar{x}}_l$ . Using Eqs. (23) and (29) leads to an equation for the (left) bath amplitudes $A_l$ in the frequency domain,

$$A_l = \frac{\omega_l}{\left(\omega_l^2 - \omega_0^2 + i\eta\omega_0\right)} \sum_k \frac{V_{l,k}}{\sqrt{\omega_l \omega_k}} A_k \tag{B11}$$

Repeating the same procedure on Eq. (11) for the molecule normal modes, and disregarding for clarity the right reservoir we get

$$\left(\omega_k^2 - \omega_0^2\right) A_k = \sum_l V_{l,k} \sqrt{\frac{\omega_k}{\omega_l}} A_l + V_{0,k} \sqrt{\frac{\omega_k}{\omega_0}} a_0^\dagger - \sum_{l,k'} \frac{V_{l,k} V_{l,k'}}{\omega_l^2} \sqrt{\frac{\omega_k}{\omega_{k'}}} A_{k'} \tag{B12}$$

Here, the driving mode ($a_0$, $\omega_0$) appears explicitly. Substitute next (B11) into (B12) yields

$$\left(\omega_k^2 - \omega_0^2\right) A_k = \sum_{l,k'} \frac{V_{l,k} V_{l,k'}}{\omega_l^2} \sqrt{\frac{\omega_k}{\omega_{k'}}} A_{k'} \left(\frac{\omega_0^2 - i\eta\omega_0}{\omega_l^2 - \omega_0^2 + i\eta\omega_0}\right) + V_{0,k} \sqrt{\frac{\omega_k}{\omega_0}} a_0^\dagger \tag{B13}$$

Note that (ignoring terms containing $1/(\omega_0 + \omega_l)$)

$$\sum_l \frac{V_{l,k} V_{l,k'}}{\omega_l^2} \left(\frac{\omega_0^2 - i\eta\omega_0}{\omega_l^2 - \omega_0^2 + i\eta\omega_0}\right) = \frac{\omega_0}{2} \sum_l \frac{V_{l,k} V_{l,k'}}{\omega_l^2} \left[-i\pi\delta(\omega_0 - \omega_l) + P\left(\frac{1}{\omega_l - \omega_0}\right)\right] \tag{B14}$$

Thus, Eq. (B13) is (after supplementing similar terms arising from the right reservoir) the same as Eq. (33).

## Appendix C. Calculation the heat flux from Eq. (38)

The classical power transfer through harmonic chain as given in [6] is calculated here between the last ($N$) atom of the molecular chain and the atoms of the right reservoir.

$$J_{0 \to r} = \frac{g_{N,r}}{m_r} \langle x_N p_r \rangle \tag{C1}$$

where 0 is a driving mode in the left reservoir and the velocity and displacement are calculated at this driving frequency $\omega_0$. The analogous quantum expression is derived from the symmetric form

$$J_{0\to r} = \frac{g_{N,r}}{2m_r}\langle x_N p_r + p_r x_N\rangle \tag{C2}$$

Transforming the coordinates into their mass weighted analogs, then expressing the local molecular coordinate in terms of molecular normal modes (Eq. (7)) yields

$$J_{0\to r} = \frac{g_{N,r}}{2m_r}\langle x_N p_r + p_r x_N\rangle = \frac{g_{N,r}}{2\sqrt{m_N m_r}}\sum_k C_{N,k}\langle \bar{x}_k \bar{p}_r + \bar{p}_r \bar{x}_k\rangle \tag{C3}$$

Using Eq. (10) and transforming (C3) into its second quantized form using (23), leads to

$$J_{0\to r} = i\sum_k \frac{V_{r,k}\omega_r}{4\sqrt{\omega_k \omega_r}} <\left(a_k^\dagger + a_k\right)\left(a_r^\dagger - a_r\right) + \left(a_r^\dagger - a_r\right)\left(a_k^\dagger + a_k\right)> \tag{C4}$$

which is the same as Eq. (41).

## Appendix D. Molecular chain model

The procedure for calculating the heat conduction through 1-dimensional model of alkane chains without invoking the harmonic approximation is represented here. The model consists of a 1 dimensional anharmonic carbon chain of length $N$ linking two reservoirs whose temperatures are denoted by $T_L$ and $T_R$. The model Hamiltonian is given by

$$\begin{aligned} H &= H_{chain} + H_{contact} \\ H_{chain} &= \sum_{i=1}^{N-1} D\left(e^{-\alpha(x_{i+1}-x_i-\tilde{x})} - 1\right)^2 + \sum_{i=1}^{N}\frac{1}{2}m\dot{x}_i^2 \\ H_{contact} &= D\left(e^{-\alpha(x_1-x_0-\tilde{x})} - 1\right)^2 + D\left(e^{-\alpha(x_{N+1}-x_N-\tilde{x})} - 1\right)^2 \end{aligned} \tag{D1}$$

where the atoms indexed by 0 and $N+1$ are the left and right reservoirs atoms respectively. In the classical simulation the positions of these atoms are taken constants, and the dynamical effect of the reservoir is represented by Langevin forces and damping terms as described below. The Morse parameters $\alpha$ and $D$ and the equilibrium bond lengths $\tilde{x}$ are taken to characterize the alkane C-C stretch motion (implying that in our classical model calculation the first bath atoms are taken to be carbons): $\tilde{x}$=1.538 Å, $D$=88 kcal/mole and $\alpha$ = 1.876 Å$^{-1}$ [43]. These force field parameters imply the spectroscopic anharmonicity coefficient (that enters in the oscillator levels $E_n = (n+1/2)\hbar\omega - (n+1/2)^2\hbar\xi\omega$)

$$\xi = \sqrt{\alpha^2 / 8D\mu} = 0.009 \tag{D2}$$

where $\mu=(1/2)m_C$ is the reduced mass. The classical equations of motions for this system are given by

$$\begin{aligned}
\ddot{x}_i &= -\frac{1}{m}\frac{\partial H}{\partial x_i}; \quad i = 2,3,...N-1 \\
\ddot{x}_1 &= -\frac{1}{m}\frac{\partial H}{\partial x_1} - \gamma_L \dot{x}_1 + F_L(t) \\
\ddot{x}_N &= -\frac{1}{m}\frac{\partial H}{\partial x_N} - \gamma_R \dot{x}_N + F_R(t)
\end{aligned} \tag{D3}$$

where $m=m_C$. In (D3) $\gamma_L$ and $\gamma_R$ are friction constants and $F_L$ and $F_R$ are fluctuating random forces that represent the effect of the thermal reservoirs. In the anharmonic calculations described in Sect. 4 we have used white reservoirs for which the damping and noise terms satisfy $\langle F_B(t)F_B(0)\rangle = 2\gamma_B m k T_B \delta(t); \; B=R,L$. The set of equations (D3) is integrated using the fourth order Runge-Kutta method, and the local heat flux is calculated from $J_i = \langle -\dot{x}_i \frac{\partial H_{i+1,i}}{\partial x_i}\rangle$ [14] where the average is done over long enough time such that the heat current is the same at all sites. Note that the expression for the heat flux is reduced in the harmonic limit into (C1), where the force constant between adjacent atoms is given by $2D\alpha^2$.

**Acknowledgements:** This work was supported by the USA-Israel Binational Science Foundation, by the Israel Academy of Science and by the Volkswagen-Stiftung under grant No. I/77 217.